# Decays of Z Boson into Pseudoscalar Meson Pair of Different Flavours

Swee-Ping Chia

*High Impact Research, University of Malaya*
*50603 Kuala Lumpur, Malaysia*

**Abstract.** We analyse the process $Z^o \to M_1 M_2$, where $M_1$ and $M_2$ are pseudoscalar mesons with quark contents of $q_1 \bar{q}$ and $q \bar{q}_2$ respectively. At the quark level, the process $Z^o \to q_1 \bar{q}_2$, where $q_1$ and $q_2$ are quarks of different flavours, receives contribution only from the Z-penguin. In order to fold the quark-level process to the hadronic process, we make the fundamental assumption that the vertex of type $Mq\bar{q}$ can be approximated by an effective constant $\gamma_5$ coupling. With this assumption, estimates are obtained for the cross-sections for the following processes: $Z^o \to K^- \pi^+$, $Z^o \to B^- K^+$.

## INTRODUCTION

Standard Model (SM) is, to a large extent, successful. It is in good agreement with the bulk of experiment data. Much efforts are now devoted to examine the finer details offered by SM. Recent discovery of the Higgs boson [1, 2] has confirmed the correctness of the Higgs mechanism, which is central to the SM.

Processes dominated by penguin diagrams have received considerable attention. Of special interest are processes involving neutral change of flavours, such as rare decay processes. Such processes are usually straightforward at the quark level. But to relate the quark-level process to the corresponding hadronic process, one has to take into consideration the QCD effects. A good understanding of how hadronic states are described in terms of quarks is, however, needed. Here, we have instead formulated a simplistic model by assuming that a pseudoscalar meson couples to quarks through an effective constant $\gamma_5$ coupling [3, 4].

The Z-penguin has attracted considerable attention in connection with the rare decays of pseudoscalar mesons and in relation to the flavour-changing decays of Z-boson [5-13]. In the present paper, I shall apply the above model to the decay of $Z^0$ into a pair of mesons of different flavours: $Z^0 \to M_1 M_2$.

## Z-PENGUIN VEREX FUNCTION FOR $Z^0 \to q_1 \bar{q}_2$

At the quark level, the process is $Z^0 \to q_1 \bar{q}_2$, where $q_1$ and $q_2$ are quarks of different flavours. This process does not arise at the tree level in the SM. The lowest order contribution to this process is from the Z-penguin.

The calculation of Higgs-penguin is performed in the 'tHooft-Feynman gauge [14]. In the 'tHooft-Feynman gauge, the diagrams in Fig.1 are accompanied by corresponding diagrams in which the internal W boson line is replaced by the unphysical charged Higgs. The vertex function calculated by summing up all these diagrams is divergent, which can be eliminated by using a renormalization scheme proposed by Chia and Chong [14, 15]. It is easily demonstrated that the counter term as calculated from the renormalization scheme above removes the divergence and yields a result identical to that would be obtained by using Ward-Takahashi Identity [13].

The proper vertex function for the Z-penguin is obtained by summing the contributions from the diagrams in Fig. 1, diagrams introduced by the unphysical charged Higgs, and the counter terms from the renormalization scheme. Putting the external quark lines on shell, we arrive at the following expression for the on-shell vertex function:

$$\Gamma_\mu = \frac{g^3}{32\pi^2 M_W^2 \cos\theta_W} \sum_j \lambda_j \left[ \left( k_\mu \slashed{k} - k^2 \gamma_\mu \right) L A_j + i\sigma_{\mu\nu} k^\nu (-m_2 L + m_1 R) B_j + \gamma_\mu L M_W^2 C_j \right]. \quad (1)$$

where $\lambda_j = V_{j2}^* V_{j1}$, $m_1$ and $m_2$ are the external quark masses, and $A_j$, $B_j$ and $C_j$ are the Z-penguin vertex form factors.

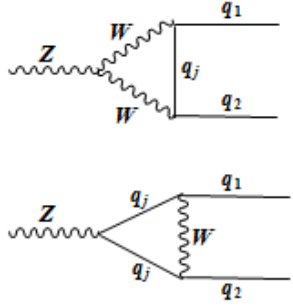

**FIGURE 1.** Fenyman diagrams employed in the calculation of Z-penguin vertex function.

The explicit expressions for the Z-penguin vertex form factors $A_j$, $B_j$ and $C_j$ are given by in Ref. 14. It is noted that these form factors develop an imaginary part whenever $k^2 > (2m_j)^2$ or $(2M_W)^2$. As we are interested in the decay of $Z^0$ boson, $k^2 = M_Z^2$. The explicit values of the vertex form factors $A_j$, $B_j$ and $C_j$ are easily calculated and are given in Table 1. Here we have used the following parameters [16]: $m_u = 2.3$ MeV, $m_c = 1.275$ GeV, $m_t = 173.5$ GeV, $M_W = 80.385$ GeV, $M_Z = 91.1876$ GeV.

**TABLE 1.** Values of Penguin Vertex Form Factors $A_j$, $B_j$, $C_j$

| j | $A_j$ | $B_j$ | $C_j$ |
|---|---|---|---|
| u | -0.366735 + i0.573962 | -0.064120 + i0.112531 | 0.380571 + i0.000000 |
| c | -0.366876 + i0.574011 | -0.064105 + i0.112469 | 0.380860 + i0.000625 |
| t | 0.168669 | 0.273896 | 3.737234 |

## THE DECAY $Z^0 \to M_1 M_2$

We next consider the decay of $Z^0$ boson into a pair of pseudoscalar mesons with different flavours. To achieve this we have to connect the Higgs penguin vertex to the external meson states. Here we make the simplifying assumption that the pseudoscalar meson couples to quark-antiquark pair through a $\gamma_5$ coupling [3, 4]. The process $Z^0 \to M_1 M_2$ therefore proceed through the loop diagram as shown in Fig.2. The loop diagram yields the following decay amplitude:

$$M = (-1) \int \frac{d^4 q_1}{(2\pi)^4} \varepsilon^\mu Tr(i\Gamma_\mu) \frac{i(\slashed{k} - \slashed{q}_1 - m_2)}{(k-q_1)^2 - m_2^2} g_2 \gamma_5 \frac{i(\slashed{q}_1 - \slashed{P}_1 - m)}{(q_1 - P_1)^2 - m^2} g_1 \gamma_5 \frac{i(\slashed{q}_1 - m_2)}{q_1^2 - m_2^2} \quad (2)$$

Here $k$ is the momentum of the $Z^0$, $P_1$ is the momentum of $M_1$, and $g_1$ and $g_2$ are the coupling constants at the $M_1$-quark and the $M_2$-quark vertices respectively. The integration over the internal momentum $q_1$ is logarithmically divergent. We introduce a cut-off momentum $\Lambda$ to tame the divergence. Assuming $\Lambda$ to be large compared to all the masses involved in the integration gives

$$M = \left( \frac{-ig^3 g_1 g_2}{1024\pi^4 M_W^2 \cos\theta_W} \right) \ln \frac{\Lambda^2}{M^2} \sum_j \lambda_j \{ (\varepsilon.k - 2\varepsilon.P_1) M_W^2 C_j - 2(\varepsilon.k k.P_1 - k^2 \varepsilon.P_1) A_j \} \quad (3)$$

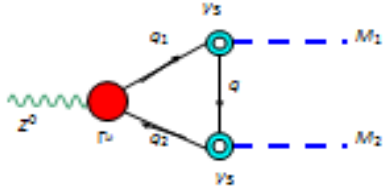

**FIGURE 2.** Diagram for the decay of $Z^0$ boson into a pair of pseudoscalar mesons with different flavours.

The decay rate $\Gamma$ is then obtained by integrating over the phase space of $\langle |M|^2 \rangle$. After some algebra, the following expression for the decay rate is obtained:

$$\Gamma = \frac{1}{6\pi} \left( \frac{g^3}{1024\pi^4 M_W^2 \cos\theta_W} \right)^2 \left( g_1 g_2 \ln \frac{\Lambda^2}{M^2} \right)^2 \frac{|\vec{P_1}|}{M_Z^2} (M_1^2 + E_1^2) \left| \sum_j \lambda_j \left( C_j + \frac{M_Z^2}{M_W^2} A_j \right) \right|^2 \quad (4)$$

The expression for $\Gamma$ involves the kinetic factor $F_{kin}$ and a dynamic form factor $F_{dyn}$ given by:

$$F_{kin} = \frac{|\vec{P_1}|}{M_Z^2} (M_1^2 + E_1^2) \quad (5)$$

$$F_{dyn} = \left| \sum_j \lambda_j \left( C_j + \frac{M_Z^2}{M_W^2} A_j \right) \right|^2 \quad (6)$$

For the kinetic factor $F_{kin}$, $|\vec{P_1}|$ and $E_1$ are given by

$$|\vec{P_1}| = \frac{1}{2M_Z} \sqrt{[M_Z^2 - (M_1 + M_2)^2][M_Z^2 - (M_1 - M_2)^2]} \quad (7)$$

$$E_1 = \frac{1}{2M_Z}(M_Z^2 + M_1^2 - M_2^2) \quad (8)$$

## DYNAMICAL FORM FACTORS $F_{DYN}$

The calculation of kinetic factor $F_{kin}$ is straightforward. To calculate the dynamic factor $F_{dyn}$, we need good estimates for the CKM matrix elements, including their relative phases. Fortunately these are available. The magnitudes of the CKM matrix elements are [16]

$$|\mathbb{V}| = \begin{pmatrix} 0.97427 & 0.22534 & 0.00351 \\ 0.22520 & 0.97344 & 0.0412 \\ 0.00867 & 0.0404 & 0.999146 \end{pmatrix}. \quad (9)$$

The relative phases can be obtained by making use of the unitary relation $\sum_j \mathbb{V}_{ij}^* \mathbb{V}_{kj} = \delta_{ik}$. With the knowledge of the CKM matrix elements, the product of CKM matrix elements, $\lambda_j$, can be calculated. Its values are as shown in Table 1 for the different processes.

**TABLE 2.** Product of CKM matrix elements $\lambda_j$

| Mode | $\lambda_u$ | $\lambda_c$ | $\lambda_t$ |
|---|---|---|---|
| $Z^0 \to K^- \pi^+$ | 0.21940 | -0.21921 –i2.6084E-4 | -1.9400E-4 +i2.6084E-4 |
| $Z^0 \to K^- B^+$ | 1.5895E-5 +i8.7588E-4 | -3.8685E-2 –i8.7588E-4 | 3.8669E-2 |

The dynamic factor $F_{dyn}$ can then be calculated from Eq. (6). Making use of the values of $\lambda_j$ from Table 2 and the penguin vertex form factors $A_j$ and $B_j$, from Table 1, we obtain the dynamical form factor $F_{dyn}$ as shown in Table 3.

**TABLE 3.** The dynamical form factor $F_{dyn}$ for the decay modes considered

| Mode | $F_{dyn}$ |
|---|---|
| $Z^0 \to K^- \pi^+$ | $0.14698 \times 10^{-5}$ |
| $Z^0 \to K^- B^+$ | $0.25286 \times 10^{-1}$ |

## DECAY RATE FOR $Z^0 \to M_1 M_2$

The calculation is now applied to the following two decay modes of $Z^0$:
(a) $Z^0 \to K^- \pi^+$
(b) $Z^0 \to B^- K^+$

The values of the parameter $g_1 g_2 \ln \Lambda^2 / M^2$ in Eq. (13) are taken from the earlier fit to the rare decays of $K$ and $B$ mesons [3, 4], as given in Table 3.

**TABLE 3.** Values of $g_1 g_2 \ln \Lambda^2/M^2$ as obtained from fit to the rare decays of K and B mesons

| Process | $g_1 g_2 \ln \Lambda^2/M^2$ |
|---|---|
| $K \to \pi e^+ e^-$ | 86.20 |
| $B \to K e^+ e^-$ | 85.44 |

Using these values of $g_1 g_2 \ln \Lambda^2/M^2$, the decay rates for the decay rates of the $Z^0$ boson are as given in Table 4.

**TABLE 4.** Decay rate $\Gamma$ and branching ratio for the different decay modes of $Z^0$

| Modes | $\Gamma$ (GeV) | Branching Ratio |
|---|---|---|
| $Z^0 \to K^- \pi^+$ | $6.625 \times 10^{-11}$ MeV | $2.67 \times 10^{-14}$ |
| $Z^0 \to K^- B^+$ | $1.138 \times 10^{-6}$ MeV | $4.58 \times 10^{-10}$ |

# CONCLUSION

The Higgs-penguin vertex is applied to the hadronic process of type $Z^0 \to M_1 M_2$. The pseudoscalar meson is assumed to couple to the quark through a $\gamma_5$ coupling. The resultant triangle diagram gives rise to a divergence which is tamed by introducing a cut-off momentum $\Lambda$. The couplings of quark to mesons $M_1$ and $M_2$ are denoted respectively by $g_1$ and $g_2$. The combinatory parameter $g_1 g_2 \ln \Lambda^2/M^2$ has been determined in an earlier fit to rare decays of K and B mesons. Using the values of $g_1 g_2 \ln \Lambda^2/M^2$ determined, the decay rates are found to be $6.625 \times 10^{-11}$ MeV and $1.138 \times 10^{-6}$ MeV respectively for the processes $Z^0 \to K^- \pi^+$ and $Z^0 \to B^- K^+$. Yhe corresponding branching ratios are $2.67 \times 10^{-14}$ and $4.58 \times 10^{-10}$ respectively.

The decay rates for $Z^0$ boson into pair of pseudoscalar mesons with different flavours are very small.

# REFERENCES

1. ATLAS Collaboration, *Phys. Lett. B* **716**, 1 (2012).
2. CMS Collaboration, *Phys. Lett. B* **716**, 30 (2012).
3. S.P. Chia, "Decays of Higgs Boson into Pseudoscalar Meson Pair of Different Flavours", in *2012 National Physics Conference, AIP Conference Proceedings Vol.* 1528, (American Institute of Physics, New York, 2012) pp. 196-200. [arXiv.1303.6392].
4. S.P. Chia, "Rare Decays of K and B Mesons Revisited", to appear in *J. Fiz. Malaysia* **34** (2013). [arXiv.1304.3195].
5. A. Axelrod, *Nucl. Phys.* **B209**, 349 (1982).
6. G. Eilam, *Phys. Rev.* D **28**, 1202 (1983).
7. M. Clements *et al.*, *Phys. Rev.* D **27**, 570 (1983).
8. V. Ganapathi *et al.*, *Phys. Rev.* D **27**, 879 (1983).
9. K.I. Hikasa, *Phys. Lett.* **148B**, 221 (1984).
10. M.J. Duncan, *Phys. Rev.* D **31**, 1139 (1985).
11. E. Ma and A. Pramudita, *Phys. Rev.* D **22**, 214 (1980).
12. T. Inami and C.S. Lim, *Prog. Theor. Phys.* **65**, 297; *ibid* 1772E (1981).
13. J.M. Soares and A. Barroso, *Phys. Rev.* D **39**, 1973 (1989).
14. N.K. Chong, "Exact Calculation of the Flavour-Changing Quark-Z Vertex", M.Sc. Thesis, University of Malaya, 1997.
15. S.P. Chia and N.K. Chong, "The Z-penguin Vertex", in *Physics at the Frontiers of the Standard Model*, ed. N.V. Hieu and J.T.T. Van (Editions Frontieres, Gif-sur-Yvette, 1996) pp. 532-535.
16. J. Beringer *et al.* (Particle Data Group), *Phys. Rev.* D **86**, 010001 (2012).